\documentclass[12pt]{article}
\usepackage{amsmath}
\usepackage{amssymb}

\usepackage{epsf,graphicx,rotating,color}
\usepackage{bm,bbm}

\newcommand{\rme}{{\rm e}}
\newcommand{\rmi}{{\rm i}}

\newcommand{\la}{\langle}
\newcommand{\ra}{\rangle}

\def\veb#1{{\bm#1}}

\newcommand{\ie}{{\it i.e.}}
\newcommand{\eg}{{\it e.g.}}

\newcommand{\mmod}{{\rm mod}}

\begin{document}

\title{Quantum computation in a Ising spin chain taking into account second
  neighbor couplings}

\author{G. V. L\' opez, T. Gorin and L. Lara\\
\small Departamento de F\'{\i}sica, Universidad de Guadalajara,\\
\small Blvd. Marcelino Garc\'{\i}a Barragan y Calzada Ol\'{\i}mpica,
        C.P. 44480 Guadalajara, Jalisco, M\'exico}

\maketitle

\begin{abstract}
We consider the realization of a quantum computer in a chain of nuclear
spins coupled by an Ising interaction. Quantum algorithms can be performed
with the help of appropriate radio-frequency pulses. In addition to the 
standard nearest-neighbor Ising coupling, we also allow for a second neighbor 
coupling. It is shown, how to apply the $2\pi k$ method in this more general 
setting, where the additional coupling eventually allows to save a few pulses.
We illustrate our results with two numerical simulations: the Shor prime 
factorization of the number $4$ and the teleportation of a qubit along a chain 
of $3$ qubits. In both cases, the optimal Rabi frequency (to suppress 
non-resonant effects) depends primarily on the strength of the second neighbor 
interaction.
\end{abstract}

PACS: 03.67.-a, 03.67.Lx, 03.65.Ta, 03.67.Dd, 03.67.Hk


\section{\label{I} Introduction}

The Ising-spin chain has been proposed in 
(Lloyd 1993, Berman et al. 1994, Lloyd 1995, Berman et al. 2001) 
as a theoretical system which allows to implement a quantum computer. 
Typically, one would think of a chain of spin-1/2 nucleons embedded into a 
solid crystal, as a possible physical system, whose dynamics may be well 
described by the Ising Hamiltonian. This system must be subjected to a magnetic
field, constant in time, with a sufficently strong variation along the spin 
chain. Additional RF-pulses (radio-frequency pulses) then allow the coherent 
control of the state of the system such that a quantum protocol can be realized
(Berman et al. 2000, Berman et al. 2001, Berman et al. 2002).
This is a special form of NMR quantum computation as described in
(Jones 2000).
Ultracold atoms in optical lattices may provide an alternative physical 
realization of a Ising-Hamiltonian (Garc\'{i}a-Ripoll and Cirac 2003).

Typically, in these system we find two types of dipolar interactions: (i)
Intrinsic dipole couplings between the nuclear spins. Those scale as distance 
$r^{-3}$ and can be cancelled by the ``magic angle'' method (Slichter 1996).
(ii) Mediated (mainly by the electrons) dipole couplings which have no clear
distance dependence. 

Up to now, this model has been developed just theoretically and hopefully the 
technological and experimental part may start in a near future. However, 
because the Hamiltonian of this system is well known, many theoretical studies 
have been made
(Berman et al. 1994, Berman et al. 2000, Berman et al. 2001, 
   Berman et al. 2002$^{(a)}$, Berman et al. 2002$^{(b)}$, 
   Berman et al. 2002$^{(c)}$, L\'{o}pez et al. 2003, Celardo et al. 2005)
which are also important for the general understanding of quantum computation. 
In this model, first neighbor Ising interaction among the nuclear spins of 
paramagnetic particles of spin one half was considered. Thus, in this paper we 
want to consider also second neighbor Ising interaction among the nuclear 
spins. The transverse coupling will be neglected since  one could expect that 
their coupling constants to be a least two or three orders of magnitude smaller 
than  the longitudinal coupling constant (Ising) for our particular 
configuration. With a register of $4$ qubits, we perform a numerical simulation 
of Shor's factorization algorithm (Shor 1994) 
of the number $4$ and teleportation (Bennett et al. 1993) 
of an arbitrary qubit in a chain of three qubits, 
allowing for an interaction between second neighbors in the system.

\section{\label{M} The model}

We consider an Ising spin chain with nearest and next-nearest neighbor 
interaction as a model for a quantum register. The spin chain is subject to
a constant external magnetic field in $z$-direction, as well as to RF-pulses
(with the magnetic field vector in the x-y plane). 
This chain is inside a strong magnetic field in the z-direction and may be
subject to RF-pulses (with the magnetic field vector in the x-y plane). The
constant magnetic field $B(z)$, which must be extremely strong, also has a 
field gradient in the $z$-direction, which allows individual addressability of 
the qubits. During an RF-pulse, the whole external field may be written as
\begin{equation}
\veb{B}= (B_0\, \cos(wt+\varphi), -B_0\, \sin(wt+\varphi), B(z)) \; ,
\end{equation}
where $B_0$, $w$ and $\varphi$ are the amplitude, the angular frequency and the 
phase of the RF-field. They are assumed to remain constant during a pulse, but
are typically chosen differently for different pulses. Without any RF-field, 
the Hamiltonian reads:
\begin{equation}
H_0= -\;{\hbar}\left( \sum_{k=1}^n w_k\; I^z_k 
 + 2J\sum_{k=1}^{n-1} I^z_k\, I^z_{k+1} + 2J'\sum_{k=1}^{n-2} I^z_k\, I^z_{k+2}
 \right) \; , 
\label{I:defH0}\end{equation}
where $w_k$ is the Larmor frequency of spin $k$. We denote with
$|0_k\rangle$ the state where the nuclear spin $k$ is parallel to the magnetic 
field and $|1_k\rangle$ where it is anti-parallel. The RF-field induces
the desired transitions between the Zeeman levels of the systems.

The structure of the Hilbert space of the spin chain is particularly 
appropriate for quantum information studies, where the basis unit of 
information is a two-level quantum system (``quantum bit'', or qubit for 
short). Any such state $\Psi=C_0|0\ra +C_1|1\ra$ can be represented with
respect to some basis stats $|0\ra$ and $|1\ra$ by two complex numbers $C_0$ 
and $C_1$ such that $|C_0|^2+|C_1|^2=1$. The $L$-tensorial product of $L$-basic
qubits form an $L$-register of $L$-qubits. In this space, we denote the
resulting product basis by 
$|\alpha\ra=|i_{L-1},\dots,i_0\rangle$ with $i_j=0,1$ for $j=0,\dots,L-1$.
A pure wave function can be expanded in this basis by
$\Psi=\sum C_\alpha|\alpha\rangle$, where $\sum_\alpha |C_\alpha|^2=1$s. For
notational convenience, we require that $\alpha= \sum_{j=0}^{L-1} i_j\, 2^j$.

The Hamiltonian $H_0$ in Eq.~(\ref{I:defH0}) is diagonal in the computational
basis (the product basis defined above):
\begin{align}
&H_0\; |\alpha_{n-1}\ldots\alpha_1\alpha_0\ra 
 = E_\alpha\; |\alpha_{n-1}\ldots\alpha_2\alpha_0\ra
\qquad I^z_k |\alpha_k\ra = \frac{(-1)^{\alpha_k}}{2}|\alpha_k\ra
 \notag\\
&E_\alpha= -\;\frac{\hbar}{2}\left( 
   \sum_{k=0}^{n-1} (-1)^{\alpha_k}\; w_k  
 + J\sum_{k=0}^{n-2} (-1)^{\alpha_k+\alpha_{k+1}}
 + J'\sum_{k=0}^{n-3} (-1)^{\alpha_k+\alpha_{k+2}}\right) \; .
\label{M:Ealpha}\end{align}
The index $\alpha$ without subscript denotes the positive integer represented
by the string $\alpha_{n-1}\ldots\alpha_1\alpha_0$ in the binary number system,
\eg: $H_0\, |101\ra = E_5\; |101\ra$. Choosing the Larmor 
frequencies $w_k$ such that $\forall k\, :\, w_k/w_{k-1} = 2$, leads to
a spectrum which (ignoring the spin-spin coupling terms) has equidistant 
levels: $E_\alpha - E_{\alpha-1} =\,$constant.

The RF-pulses are essential for any implementation of a quantum algorithm.
During such a pulse, the full Hamiltonian may be written in the form
\begin{equation}
H= H_0 + W(t)\qquad W(t)= -\frac{\hbar\,\Omega}{2}\sum_{k=1}^n 
   \left( \rme^{\rmi (wt+\varphi)}\, I^+_k + \rme^{-\rmi (wt+\varphi)}\, 
   I^-_k \right) \; ,
\label{M:RFreal}\end{equation}
where the frequency $w$, the phase-offset $\varphi$ and the Rabi frequency
$\Omega$ are free parameters ($\Omega/\gamma$ is the amplitude of the RF field, 
where $\gamma$ is the nuclear spin gyromagnetic ratio). In the computational
basis $|\alpha_{n-1}\ldots\alpha_1\alpha_0\ra$ the raising and lowering 
operators for spin $k$ can be written as: $I^+_k= |0_k\ra \la 1_k|$ and
$I^-_k= |1_k\ra \la 0_k|$. We will consider sequences of pulses, where each
pulse may vary in time. During the pulses $\varphi$ and $\Omega$ are fixed as
well as the RF frequency $w$ which is assumed to be on resonance with some
allowed transition. We assume that there are no degeneracies. Hence,
\begin{equation}
w= \frac{E_{\alpha|\alpha_k=1} -E_{\alpha|\alpha_k=0}}{\hbar} = 
   w_k + J\; \big [\, (-1)^{\alpha_{k+1}} + (-1)^{\alpha_{k-1}}\, \big ]
       + J'\; \big [\, (-1)^{\alpha_{k+2}} + (-1)^{\alpha_{k-2}}\, \big ] \; ,
\label{M:wres}\end{equation}
where it is understood that $(-1)^{\alpha_l} = 0$ if $l<0$ or $l>n$. Thus, in
the center of the spin chain (away from the borders) the $J$-coupling may lead
to frequency shifts of $\Delta w= 0$ or $\pm 2J$ ($\Delta w'= 0$ or $\pm 2J'$),
whereas at the borders the frequency shifts are $\Delta w= \pm J$
($\Delta w'= \pm J'$). The following notation is based on these observations.
The unitary evolution (in the interaction picture) during a resonant RF pulse 
is denoted by
\begin{equation}
R^{\mu,\nu}_k(\Omega\tau,\varphi)= \rme^{-\rmi H_0 \tau/\hbar}\; 
   \rme^{-\rmi H \tau/\hbar}\; \rme^{\rmi H_0 \tau/\hbar} \qquad
w= w_k + \mu J + \nu J'\; ,
\label{M:RFbase}\end{equation}
with $\mu,\nu\in[-2,-1,0,1,2]$. Note that in order that
$R^{\mu,\nu}_k(\Omega\tau,\varphi)$ is a resonant pulse, the indices 
$\mu,\nu,k$ must fulfill certain relations as discussed above (\ie\, not all 
combinations lead to resonant transitions). The wave function dynamics during
such RF-pulses is computed numerically as sketched in App.~\ref{aA}.

In the numerical simulations to follow (Secs.~\ref{F} and~\ref{T}), we have 
chosen the following parameters in units of $2\pi\times$Mhz,
\begin{equation}
\omega_0=100\ ,\ \omega_1=200\ ,\ \omega_2=400\ ,\ \omega_3=800,\ J=10\ ,\ 
J'=0.4\ ,\ \Omega=0.1\ .
\label{S:param}\end{equation}
These parameters give rise to a simple Zeeman spectrum with equidistant levels. 
For the possibly more realistic design described in 
(Berman et al. 2002$^{(b)}$), 
we would expect similar results. In our simulations, we will take advantage of 
the second neighbor interaction in order to reduce the number of pulses for the 
realization of the quantum algorithms (L\'{o}pez and Lara 2006). 
Those are Shor's factorization in Sec.~\ref{F} and teleportation in 
Sec.~\ref{T}.

\section{\label{S} Second neighbor interaction and the \boldmath 
   $2\pi k$-method}

For the considerations in the previous section, only resonant transitions have
been taken into account. That allowed trivially to describe the dynamics under
the RF-pulses analytically. However, it is possible to go beyond this simple
picture. To that end we distinguish near resonant transitions where the 
frequency differs from the resonant frequency by values of the order of $J$
or $J'$ and far resonant transitions, where the difference is of the order of
the Larmor frequencies. Then it can be shown that the Hamiltonian decomposes
into independent 2x2 matrix blocks as long as far resonant transitions
are neglected. This still allows to describe the dynamics under the RF-pulses
analytically, and in particular it allows to control and suppress the non 
(near) resonant transitions. The $2\pi k$-method (Berman et al. 2002$^{(a)}$)
and its generalization (Berman et al. 2002$^{(c)}$) 
resulted from such considerations.

Assume we perform the pulse $R_1^{0,-1}(\Omega\tau,\varphi)$ in the $4$-qubit
quantum register. This pulse induces the resonant transitions
\begin{equation}
|0001\ra \leftrightarrow |0011\ra \qquad |0100\ra \leftrightarrow |0110\ra
\end{equation}
at the frequency $w= w_1 - J'$. However, it also induces the near-resonant 
transitions at frequencies $w + \Delta$
\begin{align}
|1001\ra \leftrightarrow |1011\ra\qquad |1100\ra \leftrightarrow |1110\ra
\quad &: \quad \Delta= 2J' \notag\\
|0000\ra \leftrightarrow |0010\ra \quad &:\quad \Delta= 2J\notag\\
|0101\ra \leftrightarrow |0111\ra \quad &:\quad \Delta= -2J\notag\\
|1000\ra \leftrightarrow |1010\ra \quad &:\quad \Delta= 2J + 2J'\notag\\
|1101\ra \leftrightarrow |1111\ra \quad &:\quad \Delta= -2J + 2J'
\end{align}
In the near resonant approximation, we may write down an evolution equation
for each transition separately. To this end let $\alpha$ denote the first
state of the transition pair (where $\alpha_1=0$) and $\beta$ the second
(where $\beta_1=1$, $\forall k\ne 1\, :\, \beta_k=\alpha_k$). Then we find
for the wave function coefficients in the interaction picture (see 
App.~\ref{aA}):
\begin{equation}
\partial_t\; D_\alpha(t) = \frac{\rmi\, \Omega}{2}\; 
\rme^{\rmi (\Delta t+\varphi)}\; D_\beta(t)\qquad
\partial_t\; D_\beta(t) = \frac{\rmi\, \Omega}{2}\;
\rme^{-\rmi (\Delta t+\varphi)}\; D_\alpha(t)  \; .
\end{equation}
The solution for the evolution operator is
\begin{align}
&\begin{pmatrix} D_\alpha(t)\\ D_\beta(t)\end{pmatrix} = U(t)\; 
\begin{pmatrix} D_\alpha(0)\\ D_\beta(0)\end{pmatrix}\notag\\
&U(t)= \begin{pmatrix} \rme^{\rmi\Delta t/2} & 0\\ 0 & \rme^{-\rmi\Delta t/2}
   \end{pmatrix} \begin{pmatrix} \cos\frac{\Omega_\rme t}{2} 
      - \frac{\rmi\Delta}{\Omega_\rme}\; \sin\frac{\Omega_\rme t}{2} & 
      \frac{\rmi\Omega}{\Omega_\rme}\; \rme^{\rmi\varphi}\; 
         \sin\frac{\Omega_\rme t}{2}\\[3pt]
      \frac{\rmi\Omega}{\Omega_\rme}\; \rme^{-\rmi\varphi}\; 
         \sin\frac{\Omega_\rme t}{2} & 
      \cos\frac{\Omega_\rme t}{2} + \frac{\rmi\Delta}{\Omega_\rme} 
      \sin\frac{\Omega_\rme t}{2} \end{pmatrix} \; ,
\end{align}
where $\Omega_\rme= \sqrt{\Omega^2+\Delta^2}$.
Hence, \eg\ for a $\pi$-pulse $t=\tau=\pi/\Omega$ at the end of the pulse,
and we may tune $\Omega$ such that any near resonant transition is switched
off. This only requires that
\begin{equation}
\frac{\Omega_\rme \tau}{2} = \frac{\pi\, \Omega_\rme}{2\, \Omega}
 = \frac{\pi}{2}\sqrt{1+\frac{\Delta^2}{\Omega^2}} = k\pi 
\quad\Leftrightarrow\quad
\sqrt{1+\frac{\Delta^2}{\Omega^2}} = 2\, k \; .
\end{equation}
The $2\pi k$-method as presented here has two
shortcomings. If there are several near-resonant transitions with different
frequency shifts $\Delta$ it will be impossible, in general, to eliminate all
of them. Second, any near resonant transition also implies a phase rotation,
which cannot be corrected by the method. In principle some of those 
shortcomings can be overcome at the expense of more complex pulse sequences,
as shown in (Berman et al. 2002c). 

For the numerical simulations in Secs.~\ref{F} and~\ref{T} we define the 
optimal Rabi frequency to eliminate a near-resonant transition with shift 
$\Delta$ during a $\pi$-pulse as
\begin{equation}
\Omega_\Delta^{(k)}= \frac{|\Delta|}{\sqrt{4k^2-1}} \; .
\label{S:omegak}\end{equation}
Note that for a $\pi/2$-pulse the corresponding optimal Rabi frequency is
given by the same equation, but with $k$ replaced by $2k$.

\section{\label{F} Simulations of Shor's factorization algorithm of number 
   four}

An experimental realization of Shor's factorization algorithm has been 
demonstrated recently in (Vandersypen et al. 2001) 
using nuclear magnetic resonance. Following Shor's approach (Shor 1994)
for factorizing an integer number 
$N$, one selects a $L+M$-register of the form $|x;y\rangle$, where 
$|x\rangle$ is the input register of length $L$, and $|y\rangle$ is the
evaluation register of length $M$. The $y$-register is used to store the 
values of the periodic function $y(x)=q^x(\mmod~N)$, where the integer $q$ is 
chosen co-prime to $N$ (\ie\ the greatest common divisor $\gcd(q,N)=1$). The 
algorithm is divided into three parts (Nielsen and Chuanf 2000, chapter 5).
First, one creates the uniform superposition of all basis states in the 
$x$-register. Second, one chooses some co-prime number $q$ and computes the 
function $y(x)=q^x(\mmod~N)$ in the $y$-register. Third, one applies the 
inverse discrete Fourier transform to the $x$-register. After these steps, one 
measures the state in the $x$-register, which provides the information on the 
factors of $N$. For more details, we refer the interested reader to 
(Nielsen and Chuang 2000).  

For factorizing the number $N=4$, two qubits in each register are sufficient
($L=M=2$). The only co-prime number of $N$ is $q=3$, and thus the period of the 
function $y(x)=3^x(\mmod~4)$ gives the factors of $N$. In the present case this 
period is $T=2$. Applying the procedure above to the initial state 
$\Psi_0= |00;00\ra$, one obtains after each of the three steps:
\begin{align}
\Psi^{({\rm ideal})}_1 
 &= \frac{1}{2}\big (\, |00;00\ra + |01;00\ra + |10;00\ra + |11;00\ra\, \big )
    \\
\Psi^{({\rm ideal})}_2 
 &= \frac{1}{2}\big (\, |00;01\ra + |01;11\ra + |10;01\ra + |11;11\ra\, \big )
   \\
\Psi^{({\rm ideal})}_3 
 &= \frac{1}{2}\big (\, |00;01\ra + |00;11\ra + |10;01\ra + |10;11\ra\, \big )
   \; .
\label{F:Psi3}\end{align}
The measurement on the $x$-register give us the states $|00\ra$ or $|10\ra$ 
($x=0$ or $x=2$), which implies that the period of the function $y(x)$ is
$T=2$, as expected.

The following computation is performed in the interaction picture.
Using the parameters from Eq.~(\ref{S:param}) and starting with the ground 
state of the system, $\Psi_0=|0000\ra$, we create $\Psi_1$ with the help of 
three $\pi/2$-pulses:
\begin{equation}
\Psi_1= R_3^{-1,1}(\pi/2,\pi/2)\; R_3^{1,1}(\pi/2,\pi/2)\;
   R_2^{2,1}(\pi/2,\pi/2)\; \Psi_0 \; .
\label{F:sequ1}\end{equation}
The evaluation of the function $y(x)=3^x(\mmod~4)$ in the $y$-register is 
carried out with four $\pi$-pulses:
\begin{equation}
\Psi_2= R_1^{-2,-1}(\pi,\pi/2)\; R_1^{-2,1}(\pi,\pi/2)\;
   R_0^{1,-1}(\pi,\pi/2)\; R_0^{1,1}(\pi,\pi/2)\; \Psi_1 \; .
\end{equation}
Finally, the discrete Fourier transformation in the $x$-register is obtained 
through five $\pi$-pulses:
\begin{equation}
\Psi_3= R_2^{-2,-1}(\pi,\pi/2)\; R_0^{-1,-1}(\pi,-\pi/2)\;
   R_0^{-1,1}(\pi,\pi/2)\; R_2^{0,1}(\pi,-\pi/2)\; R_0^{-1,-1}(\pi,-\pi/2)\; 
   \Psi_2 \; .
\label{F:sequ3}\end{equation}
The final measurement consists in tracing out the $y$-register, and measuring
the probabilities of finding the $x$-register in any of the possible basis 
states. In the present case, we would get
\begin{equation}
\la 00|\varrho_x|00\ra = \la 10|\varrho_x|10\ra = \frac{1}{2}\qquad
\la 01|\varrho_x|01\ra = \la 11|\varrho_x|11\ra = 0 \; ,
\end{equation}
where $\varrho_x= {\rm tr}_y |\Psi_3\ra\la\Psi_3|$. This yields the expected 
period $T=2$.

\begin{figure}
\input{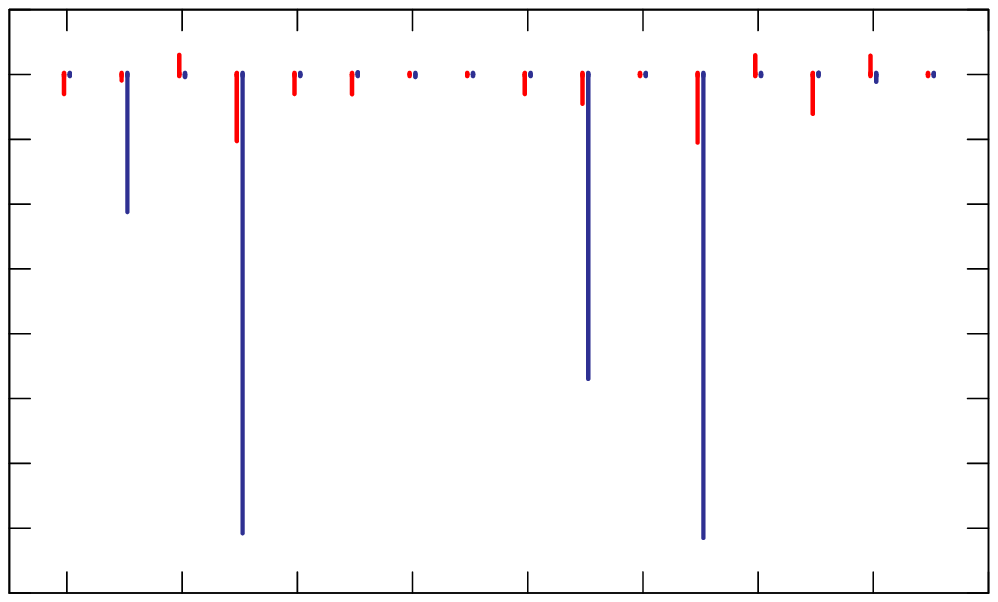}
\caption{The difference in the expansion coefficients between the final state
$\Psi_3$ obtained from the evolution with the full Hamiltonian and the
ideal final state $\Psi^{({\rm ideal})}_3$. Red impulses (towards the left) 
show the real part, blue impulses (towards the right) the imaginary part. 
The difference in the expansion coefficients is plotted versus
$\alpha$, which labels the basis states.}
\label{f:X1}\end{figure}

In what follows, we simulate the pulse sequence for Shor's algorithm with our
model Hamiltonian of Eq.~(\ref{M:RFreal}). The wave-packet evolution is carried
out in the interaction picture using the parameters given in 
Eq.~(\ref{S:param}). In Fig.~\ref{f:X1} we show the difference between the
final state $\Psi_3$, obtained from the evolution with the full Hamiltonian 
$H(t)$, and the ideal final state $\Psi^{({\rm ideal})}_3$. The latter is
obtained from an evolution where only the resonant transitions have been taken
into account. As deviation, we plot the real (red impulses) and imaginary 
(blue impulses) parts of the difference between the respective expansion 
coefficients $D_\alpha$ and $D^{({\rm ideal})}_\alpha$ as a function of 
$\alpha$, which denotes the basis states, as explained below
Eq.~(\ref{M:Ealpha}). We can see that by far the largest error consist in an
accumulation of residual phases in the expansion coefficients of those basis
stats which are expected to be populated in $\Psi^{({\rm ideal})}_3$, 
\ie\ $|0001\ra$, $|0011\ra$, $|1001\ra$, and $|1011\ra$. This can be deduced
from the large imaginary parts errors, visible in Fig.~\ref{f:X1}.
However, the probabilities of the states remain within their right value.

\begin{figure}
\input{figX3.tex}
\caption{The fidelity, Eq.~(\ref{F:defF}), for the final state resulting from 
the factorization algorithm as a function of the Rabi frequency $\Omega$. The
blue arrows show the optimal Rabi frequencies $\Omega^{(k)}_\Delta$ according 
to Eq.~(\ref{S:omegak}), for $\Delta=2J'$ and $k=1,2,3$, and $4$ (from right
to left).}
\label{f:X2}\end{figure}


In order to quantify the deviation of the real evolution with $H(t)$ from the 
ideal one, where only resonant transitions are taken into account, we introduce
a figure of merit, the fidelity 
(Peres 1984, Schumacher 1995, Gorin et al. 2006)  
\begin{equation}
F= |\la\Psi^{({\rm ideal})}|\Psi\ra|^2 \; ,
\label{F:defF}\end{equation}
which gives the probability to find the system in the state 
$\Psi^{({\rm ideal})}$ even though it really is in the state $\Psi$.
As we have seen in Sec.~\ref{S} that one can influence the accuracy of the
implementation of the quantum protocol by appropriately choosing the Rabi 
frequency (the strength of the RF field) we will first investigate the 
dependence of the fidelity 
$F_{\rm fi}= |\la\Psi_3^{({\rm ideal})}|\Psi_3\ra|^2$ on the Rabi frequency.
The result is shown in Fig.~\ref{f:X2}, where $F_{\rm fi}$ is plotted versus
$\Omega$ in the interval $(0.08,0.48)$. The blue arrows show the positions of
the optimal Rabi frequencies $\Omega^{(k)}_\Delta$, according to 
Eq.~(\ref{S:omegak}) for $\Delta= 2J'$ and $k=1,2,3$, and $4$ (from right to 
left). The fidelity clearly behaves as expected. We find maxima for those
values, where the $2\pi\, k$-relation is fulfilled. The resulting curve is
rather smooth at small Rabi frequencies, while additional fast oscillations
appear near $\Omega\approx 0.3$. The additional structures are due to the
transitions where the detuning is of the order of $2J$. Since $J\gg J'$ these
transitions have nearly no effect for small Rabi frequencies, but there effect
becomes more and more significant as $\Omega$ increases. In practice, one
would have to find a compromise between a sufficiently fast evolution (for that
one needs large Rabi frequencies) and a sufficiently high fidelity, which can
be achieved more easily at small Rabi frequencies.


\begin{figure}
\input{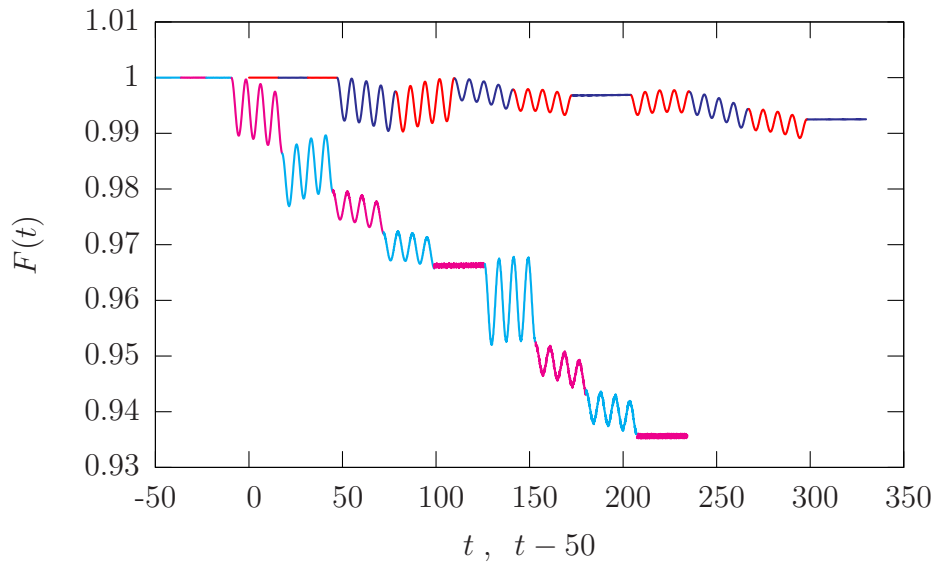}
\caption{The fidelity, Eq.~(\ref{F:defF}), during the quantum factorization
algorithm as a function of time $t$. We show two cases: 
$\Omega= \Omega^{(2)}_\Delta= 0.1$ for $\Delta= 2J'$ (red and blue lines), and 
$\Omega= 0.116$ (light blue and pink lines), where the fidelity of the final 
state has a minimum (see Fig.~\ref{f:X2}). For each case, we used alternating 
colors to distinguish the individual pulses in the pulse sequence 
Eqs.~(\ref{F:sequ1}--\ref{F:sequ3}). For better visibility, the curve for 
$\Omega= 0.116$ is shifted to the left by $50$ time units.}
\label{f:X4}\end{figure}


In order to study the mechanism behind the $2\pi\, k$-method in more detail, we
record the (loss of) fidelity during the execution of Shor's quantum
algorithm as a function of time. The result is shown for two cases,
$\Omega= \Omega^{(2)}_\Delta= 0.1$ for $\Delta= 2J'$, and $\Omega= 0.116$ in
Fig.~\ref{f:X4}. In the first case, the Rabi frequency is a optimal one,
according to Eq.~(\ref{S:omegak}), whereas in the second case, the Rabi
frequency is such that $F_{\rm fi}$, as plotted in Fig.~\ref{f:X2} has a
minimum. In both cases, we see the expected Rabi oscillations (in $F(t)$)
during the majority of the $\pi$-pulses. In the first case, the 
oscillations are completed during the $\pi$-pulses, such that $F(t)$ at the
end of a $\pi$-pulse is almost returns to the value it had at the beginning
of the pulse. By contrast, in the second case, the $\pi$-pulses stop when
$F(t)$ is almost exactly in a minimum of the near-resonant Rabi oscillations.
In the graph of the second case (light blue and pink lines) we can see that for
some pulses (\eg\ the $8$'th and the $11$'th) there are apparently no 
oscillations, only a striking broadening of the curve for $F(t)$. During those
pulses, the near-resonant transitions with $\Delta= 2J'$ involve states which
almost unoccupied, such that the transitions cannot really take place. In
those instances we have a chance to see the effects of transitions with 
$\Delta$ being of the order of $2J$. The resulting oscillations in $F(t)$ 
are too fast to be resolved, such that the only visible effect is a broadening
of the curve. This also demonstrates that the transitions with detuning of the
order of $2J$ can be neglected at small Rabi frequencies. That explains the
smooth behavior of $F_{\rm fi}(\Omega)$, depicted in Fig.~\ref{f:X2}.

\section{\label{T} Quantum teleportation on 3-qubits}

The basic idea of quantum teleportation (Bennett et al. 1993) 
is that Alice (left end 
qubit in our chain of three qubits) and Bob (the other end qubit) share two
qubits which are in a maximally entangled (Bell) state. Here, we shall follow
the prescription of (Nielsen and Chuang 2000). 
\begin{equation}
\Phi_\rme= \frac{1}{\sqrt{2}}\; \big (\, 
   |0_{\rm A}00_{\rm B}\ra + |1_{\rm A}01_{\rm B}\ra\, \big ) \; .
\end{equation}
We adjoin to Alice the arbitrary state
\begin{equation}
\Phi_{\rm x}= C_0^{\rm x}|0\ra+C_1^{\rm x}|1\ra\ ,
\end{equation}
to be ``teleported'' to Bob. This results in the quantum state 
($\Phi_1=\Phi_{\rm x}\otimes\Phi_\rme$) of the whole system 
\begin{equation}
\Phi^{({\rm ideal})}_1=\frac{1}{\sqrt{2}}\; \big (\, C_0^{\rm x}\; |0000\ra
   + C_0^{\rm x}\; |0101\ra + C_1^{\rm x}\; |1000\ra 
   + C_1^{\rm x}\; |1101\ra\, \big )\ .
\end{equation}
Then, we apply a controlled-Not (CN) operation between the added qubit and
that of Alice
\begin{equation}
\widehat{\rm CN}_{32}\; |i_3,i_2,i_1,i_0\ra
 = |i_3, i_2\oplus i_3, i_2, i_0\ra\ ,
\end{equation}
where $i_2\oplus i_3=(i_2+i_3) \mmod~2$. This results in the state 
\begin{equation}
\Phi^{({\rm ideal})}_2= \frac{1}{\sqrt{2}}\; \big (\, C_0^{\rm x}\; |0000\ra 
   + C_0^{\rm x}\; |0101\ra + C_1^{\rm x}\; |1100\ra
   + C_1^{\rm x}\; |1001\ra\, \big )\ .
\end{equation}
In (Nielsen and Chuang 2000)  
a final Hadamard gate is applied to the added qubit. It
allows Alice to measure the two qubits in the computational basis instead of 
the Bell basis; see (Bennett et al. 1993).  
In our implementation, we replace the
Hadamard gate with a different $\pi/2$-qubit rotation $\widehat{\rm A}$, which 
however achieves the same goal:
\begin{equation}
\widehat{\rm A}_3\; \begin{cases} |0000\ra\\ |1000\ra\end{cases} 
 = \frac{1}{\sqrt{2}}\;
      \begin{cases} |0000\ra - |1000\ra\\ |0000\ra + |1000\ra \end{cases}\ .
\end{equation}
On the Bloch-sphere, this corresponds to a rotation about the angle $-\pi/2$
around the $y$-axis. After some rearrangements, the final state 
$\Phi^{({\rm ideal})}_3= \widehat{\rm A}_3\; \Phi^{({\rm ideal})}_2$ reads:
\begin{align}
\Phi^{({\rm ideal})}_3 
 &= \frac{1}{2}\left\{ |00\ra\otimes|0\ra\big (\, C_0^{\rm x}\; |0\ra
   + C_1^{\rm x}\; |1\ra\, \big ) 
   + |01\ra\otimes|0\ra\otimes\big (\, C_0^{\rm x}\; |1\ra + C_1^{\rm x}\;
      |0\ra\, \big )\right. \notag\\
&\qquad + \left. |10\ra\otimes|0\ra\otimes\big (\, C_1^{\rm x}\; |1\ra
   - C_0^{\rm x}\; |0\ra\, \big )
   + |11\ra\otimes|0\ra\otimes\big (\, C_1^{\rm x}\; |0\ra 
   - C_0^{\rm x}\; |1\ra\, \big )\right\}\ .
\label{T:Phi3ideal}\end{align}
When Alice measures both of her qubits, there are four possible cases:
$|00\ra$, $|01\ra$, $|10\ra$, and $|11\ra$. For each case, Bob will get the 
original state $\Phi_{\rm x}$, provided he applies the proper operation to
his qubits. The operation he has to choose depends on the outcome of Alice's
measurement as follows:
\begin{equation}
\begin{array}{c|c}
\text{Alice's result} & \text{Bob's operation}\\
\hline
|00\ra & \widehat{\rm id} \\
|01\ra & \widehat{\rm N}  \\
|10\ra & -\sigma_z        \\
|11\ra & \widehat{\rm N}\, \sigma_z
\end{array}\qquad\text{where}\quad
\begin{array}{rl}
\widehat{\rm id} &: |0\ra \to |0\ra\, ,\; |1\ra \to |1\ra\\
\widehat{\rm N} &: |0\ra \to |1\ra\, ,\; |1\ra \to |0\ra\\
\sigma_z &: |0\ra \to |0\ra\, ,\; |1\ra \to -|1\ra
\end{array}\ .
\end{equation}
Note that neither Alice nor Bob need to know the state $\Phi_{\rm x}$ in order
to perform the transfer to Bob's qubit. In fact if any of the two ``knew'' 
something about the state, the state transfer would not work.

In order to implement the quantum teleportation scheme on our Ising spin chain
quantum computer, we start with a spin chain of $4$ spins ($0,\ldots 3$),
where spin $3$ is in the unknown state $\Phi_{\rm x}$.
\begin{equation}
\Phi_0= C_0^{\rm x}\; |0000\ra + C_1^{\rm x}\; |1000\ra \; .
\end{equation}
The first step consists in generating an entangled (Bell) state between qubit 
$2$ (Alice) and qubit $0$ (Bob). This is obtained with the following three 
pulses:
\begin{equation}
\Phi_1= R_0^{1,-1}(\pi,-\pi/2)\; R_2^{0,1}(\pi/2,-\pi/2)\;
   R_2^{2,1}(\pi/2,-\pi/2)\; \Phi_0 \; .
\label{T:sequ1}\end{equation}
Then the controlled-not operation $\widehat{\rm CN}_{32}$ is applied through
\begin{equation}
\Phi_2= R_2^{0,-1}(\pi,-\pi/2)\; R_2^{0,1}(\pi,\pi/2)\; \Phi_1 \; .
\end{equation}
The final state, given in Eq.~(\ref{T:Phi3ideal}), is obtained after applying 
the $\widehat{A}_3$ to qubit $3$ via the following two pulses:
\begin{equation}
\Phi_3= R_3^{-1,1}(\pi/2,-\pi/2)\; R_3^{1,1}(\pi/2,-\pi/2)\; \Phi_2 \; .
\label{T:sequ3}\end{equation}
As an example, we choose $C_0^{\rm x}= 1/3$ and $C_1^{\rm x}= \sqrt{8}/3$ for
the coefficients of the unknown state.

\begin{figure}
\input{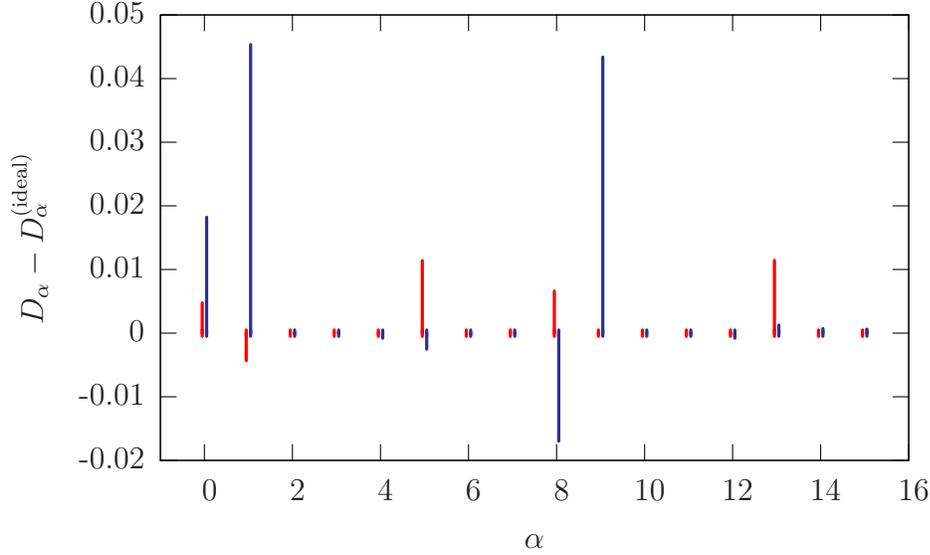}
\caption{The difference in the expansion coefficients between the final state
$\Phi_3$ obtained from the evolution with the full Hamiltonian and the
ideal final state $\Phi^{({\rm ideal})}_3$. Red impulses (towards the left) 
show the real part, blue impulses (towards the right) the imaginary part.
The difference in the expansion coefficients is plotted versus
$\alpha$, which labels the basis states.}
\label{f:X6}\end{figure}

Again, we simulate the pulse sequence with our model Hamiltonian of 
Eq.~(\ref{M:RFreal}) using the parameters of Eq.~(\ref{S:param}), just as in 
the previous section. Fig.~\ref{f:X6} shows the difference between the final 
state $\Phi_3$, obtained from the simulation and the ideal final state 
$\Phi^{({\rm ideal})}_3$, given in Eq.~(\ref{T:Phi3ideal}). The figure is 
produced exactly in the same way, as Fig.~\ref{f:X1} for the case of Shor 
factorization. Here, we find somewhat smaller deviations than in 
Fig.~\ref{f:X1}, which may be simply explained by the number of pulses for the 
teleportation protocol being smaller. 

\begin{figure}
\input{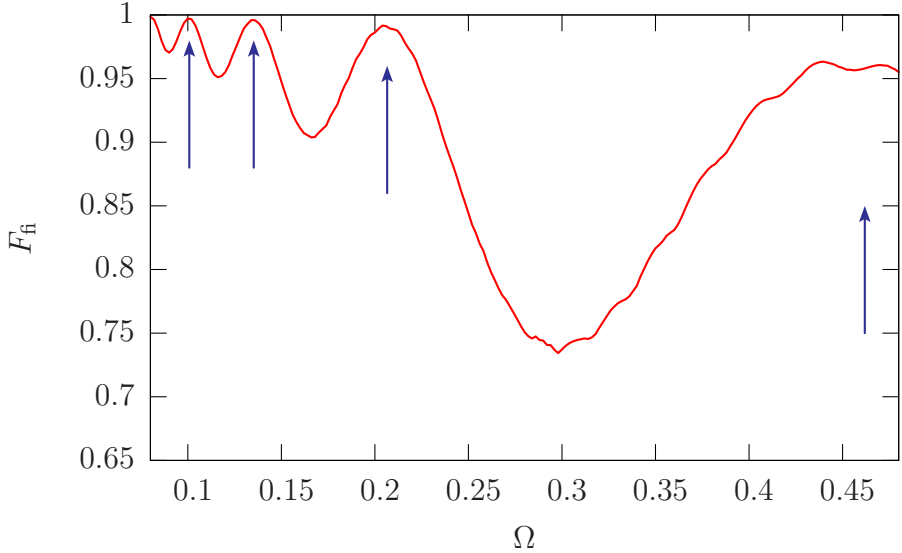}
\caption{The fidelity, Eq.~(\ref{F:defF}), for the final state resulting from 
the teleportation algorithm as a function of the Rabi frequency $\Omega$. The
blue arrows show the optimal Rabi frequencies $\Omega^{(k)}_\Delta$ according 
to Eq.~(\ref{S:omegak}), for $\Delta=2J'$ and $k=1,2,3$, and $4$ (from right
to left).}
\label{f:X7}\end{figure}

In Fig.~\ref{f:X7} we show the fidelity 
$F_{\rm fi}= |\la\Phi_3^{({\rm ideal})}|\Phi_3\ra|^2$ as a function of the
Rabi frequency $\Omega$. Again, Fig.~\ref{f:X7} is the precise analog of
Fig.~\ref{f:X2}, where the same quantity is plotted for the Shor factorization
protocol. We again find the expected maxima at those points, where the
$2\pi\, k$-relation, Eq.~(\ref{S:omegak}), is fulfilled. At least for 
sufficiently small frequencies, there are no noticeable effects of the nearest
neighbor interaction, which implies much larger detuning $\Delta\sim 2J$. 
The present protocol contains $\pi$-pulses, as well as $\pi/2$-pulses. 
However, note that for $\pi/2$-pulses, the $2\pi\, k$-relation is fulfilled
for even $k$, only -- see discussion below Eq.~(\ref{S:omegak}). On 
Fig.~\ref{f:X7}, the corresponding frequencies are marked by the second and the
forth arrow, respectively (counting from the right). The data does not show
any modulation with respect to $k$ being odd or even. This is a clear 
signature that the fidelity of the present protocol is largely determined by
the $\pi$-pulses alone.

\begin{figure}
\input{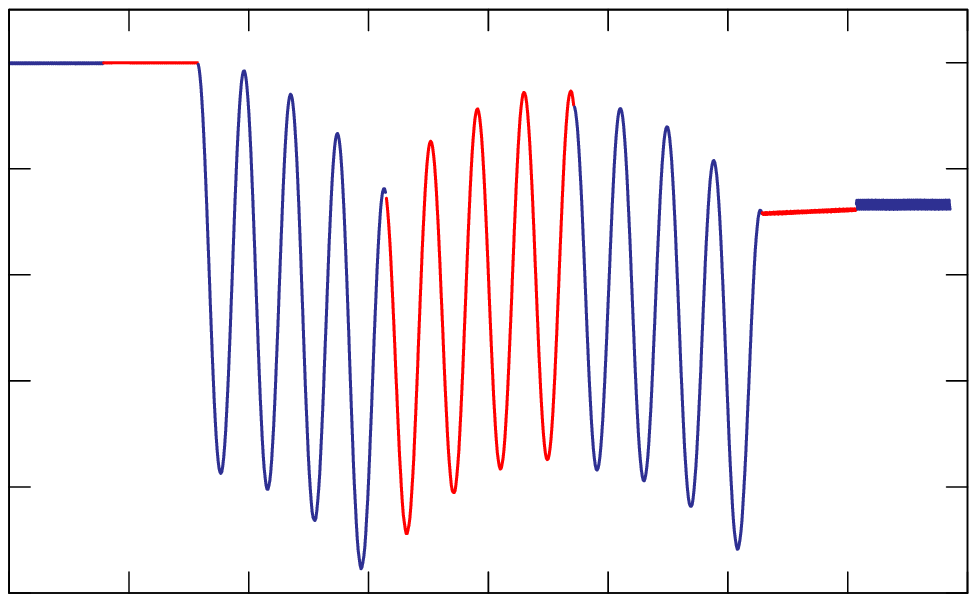}
\caption{The fidelity, Eq.~(\ref{F:defF}), during the quantum factorization
algorithm as a function of time $t$ for $\Omega= \Omega^{(2)}_\Delta= 0.1$,
where $\Delta= 2J'$ (red and blue lines). We used alternating colors to 
distinguish the individual pulses in the pulse sequence 
Eqs.~(\ref{T:sequ1}--\ref{T:sequ3}).}
\label{f:X8}\end{figure}

Fig.~\ref{f:X8} shows the (loss of) fidelity during the execution of the
teleportation protocol as a function of time. In the present case, we choose
$\Omega= \Omega^{(2)}_\Delta= 0.1$ for $\Delta= 2J'$. Thus, the Rabi frequency
fulfills the $2\pi\, k$-relation, Eq.~(\ref{S:omegak}), such that the final
fidelity is maximal, locally (see Fig.~\ref{f:X7}). In Fig.~\ref{f:X8} we can
distinguish the $\pi/2$ pulses (two on each side) surrounding the three 
$\pi$-pulses in the middle of the pulse sequence. Oscillations are visible
only during the central $\pi$-pulses, not during any of the $\pi/2$-pulses.
Note, however that during the last $\pi/2$-pulse we find very fast oscillations
with a relatively small amplitude, which must come from detuning of the order
of $\Delta\sim 2J$. The behavior of $F(t)$ clearly demonstrates that the loss
of fidelity occurs almost exclusively during the execution of the $\pi$-pulses.



\section{\label{C} Conclusion}

For a quantum computer realized with a one-dimensional chain of nuclear spins 
(one half), we have studied the effect of an additional second 
neighbor Ising interaction. This allows greater flexibility in the choice of
appropriate pulse sequences for the implementation of a given quantum 
algorithm. We have found that after adapting the $2\pi k$-method 
to this new situation, the desired gate operations can be realized with high
fidelity. We have illustrated our results with two case studies: Shor's 
quantum algorithm for the factorization of an integer number, and another
algorithm which allows the teleportation of a qubit across the spin chain. In
both cases, we have studied the fidelity of the whole algorithm as a function
of the Rabi frequency, as well as the decay of fidelity during the algorithm
as a function of time. In spite of the fact that both algorithms contain 
$\pi$- and $\pi/2$-pulses, we found that only the $\pi$-pulses were responsible 
for the decay of fidelity. We found also that as long as the nearest-neighbor 
interaction $J$ is much larger than that of the next-nearest neighbor 
interaction $J'$, the (loss of) fidelity is dominated by the latter. In other 
words, if the second neighbor interaction in a chain of nuclear spins is not
entirely negligible, it will typically dominate the non-resonant effects.

\section*{Acknowledgements}

This work was supported by SEP under the contract PROMEP/103.5/04/1911 and the 
University of Guadalajara.

\begin{appendix}

\section{\label{aA} Numerical solution of the Schr\" odinger equation}

We consider the time evolution of the quantum state $\Psi(t)$ of the Ising
spin chain during a RF pulse as defined in Eq.~(\ref{M:RFreal}). 
The Schr\" odinger equation reads:
\begin{equation}
\rmi\hbar\,\partial_t\; \Psi(t) = H(t)\; \Psi(t) \qquad
H(t)= H_0 - \frac{\hbar\Omega}{2} \sum_{k=1}^n 
   \left( \rme^{\rmi (wt+\varphi)}\, I^+_k + \rme^{-\rmi (wt+\varphi)}\, 
   I^-_k \right) \; ,
\end{equation}
where $\Psi(0)$ is the quantum state at the beginning of the pulse, and 
$\Psi(\tau)$ is the state at the end ($\tau$ is the duration of the pulse).
The unperturbed Hamiltonian $H_0$ is given in Eq.~(\ref{I:defH0}). Expanding 
$\Psi(t)$ in the basis of the time-independent part $H_0$, we arrive at the 
matrix form of the Schr\" odinger equation:
\begin{equation}
\rmi\partial_t\, C_\alpha(t) = \frac{E_\alpha}{\hbar}\; C_\alpha(t)
   - \frac{\Omega}{2} \sum_{k=0}^{n-1} 
      \begin{cases} 
         \rme^{\rmi (wt+\varphi)}\, C_{\alpha|\alpha_k=1}(t) &: \alpha_k=0\\
         \rme^{-\rmi (wt+\varphi)}\, C_{\alpha|\alpha_k=0}(t) &: \alpha_k=1
      \end{cases} \; .
\end{equation}
In order to obtain the dynamics in the interaction picture, we set
\begin{equation}
C_\alpha= D_\alpha\; \rme^{-\rmi E_\alpha t/\hbar} \quad\Rightarrow\quad
\rmi\partial_t\, D_\alpha(t) = - \frac{\Omega}{2} \sum_{k=0}^{n-1}
   \begin{cases} 
      \rme^{\rmi (\Delta_k t+\varphi)} D_{\alpha|\alpha_k=1}(t) &: \alpha_k=0\\
      \rme^{-\rmi (\Delta_k t+\varphi)} D_{\alpha|\alpha_k=0}(t) &: \alpha_k=1
   \end{cases} \; ,
\end{equation}
where 
\begin{equation}
\Delta_k= w- w_k 
   - J\; \big [\, (-1)^{\alpha_{k+1}} + (-1)^{\alpha_{k-1}}\, \big ]
   - J'\; \big [\, (-1)^{\alpha_{k+2}} + (-1)^{\alpha_{k-2}}\, \big ] \; ,
\end{equation}
as obtained from Eq.~(\ref{M:wres}). Here, it is understood that 
$(-1)^{\alpha_l} = 0$ if $l<0$ or $l>n$. In order to evolve the wave packet
numerically, it is convenient to rewrite the system of differential equations
in terms of real variables. To that end, we introduce the real variables
$X_\alpha(t)$ and $Y_\alpha(t)$ such that $D_\alpha= X_\alpha+\rmi Y_\alpha$.
Then we find:
\begin{align}
\partial_t\, X_\alpha(t) &= \frac{\Omega}{2}\sum_{k=0}^{n-1}
   \begin{cases}
    - \sin(\Delta_k t+\varphi)\; X_{\alpha|\alpha_k=1}(t)
    - \cos(\Delta_k t+\varphi)\; Y_{\alpha|\alpha_k=1}(t) &: \alpha_k=0\\
      \sin(\Delta_k t+\varphi)\; X_{\alpha|\alpha_k=0}(t)
    - \cos(\Delta_k t+\varphi)\; Y_{\alpha|\alpha_k=0}(t) &: \alpha_k=1
   \end{cases} \notag\\
\partial_t\, Y_\alpha(t) &= \frac{\Omega}{2}\sum_{k=0}^{n-1}
   \begin{cases}
      \cos(\Delta_k t+\varphi)\; X_{\alpha|\alpha_k=1}(t)
    - \sin(\Delta_k t+\varphi)\; Y_{\alpha|\alpha_k=1}(t) &: \alpha_k=0\\
      \cos(\Delta_k t+\varphi)\; X_{\alpha|\alpha_k=0}(t)
    + \sin(\Delta_k t+\varphi)\; Y_{\alpha|\alpha_k=0}(t) &: \alpha_k=1
   \end{cases} \; .
\end{align}

\end{appendix}

\bibliographystyle{unsrt}
\bibliography{amol,deco,ranh,semic,stas,books,qcom,echo}

\end{document}